# Cartesian dictionary-based native $T_1$ and $T_2$ mapping of the myocardium

**Author:**

Markus Henningsson[1,2]

1. Division of Diagnostics and Specialist Medicine, Department of Health, Medicine and Caring Sciences (HMV), Linköping University, Linköping, Sweden
2. Center for Medical Image Science and Visualization (CMIV), Linköping University, Linköping, Sweden

**Correspondence:** Markus Henningsson, Division of Cardiovascular Medicine, Department of Medical and Health Sciences, Linköping University, Linköping, Sweden

Email: markus.henningsson@liu.se

Word count: 5018

# Abstract

**Purpose:** To implement and evaluate a new dictionary-based technique for native myocardial $T_1$ and $T_2$ mapping using Cartesian sampling.

**Methods:** The proposed technique (Multimapping) consisted of single-shot Cartesian image acquisitions in 10 consecutive cardiac cycles, with inversion pulses in cycle 1 and 5, and $T_2$ preparation (TE: 30ms, 50ms and 70ms) in cycles 8-10. Multimapping was simulated for different $T_1$ and $T_2$, where entries corresponding to the k-space centers were matched to acquired data. Experiments were performed in a phantom, 16 healthy subjects and three patients with cardiovascular disease.

**Results:** Multimapping phantom measurements showed good agreement with reference values for both $T_1$ and $T_2$, with no discernable heart-rate dependency for $T_1$ and $T_2$ within the range of myocardium. In vivo mean $T_1$ in healthy subjects was significantly higher using Multimapping ($T_1$=1114±14ms) compared to the reference ($T_1$=991±26ms) ($p$<0.01). Mean Multimapping $T_2$ (47.1±1.3ms) and $T_2$ spatial variability (5.8±1.0ms) was significantly lower compared to the reference ($T_2$=54.7±2.2ms, $p$<0.001; spatial variability=8.4±2.0ms, $p$<0.01). Increased $T_1$ and $T_2$ was detected in all patients using Multimapping.

**Conclusions:** Multimapping allows for simultaneous native myocardial $T_1$ and $T_2$ mapping with a conventional Cartesian trajectory, demonstrating promising in vivo image quality and parameter quantification results.

## Introduction

In the last decade, there has been tremendous interest in developing and clinically applying MRI techniques for quantitatively mapping relaxation times in the myocardium (1–3). Myocardial $T_1$ and $T_2$ mapping can be used to discriminate between a range of different cardiomyopathies (4). $T_1$ mapping can be used to detect diffuse and focal fibrosis, and can also be employed to calculate extracellular volume fraction which is altered in many cardiomyopathies, while $T_2$ mapping is most commonly used to assess myocardial edema and inflammation (5–7). For some diseases, a combination of $T_1$ or $T_2$ mapping provides the most diagnostic power, for example to detect myocarditis (8,9).

The most clinically used quantitative techniques enable $T_1$ or $T_2$ mapping as separate scans. For $T_1$ mapping, the modified Look-Locker inversion recovery (MOLLI) has been the most widely used due to its relatively high precision (10,11). There is less consensus regarding the optimal $T_2$ mapping technique, and methods using either dark-blood fast spin echo or bright-blood $T_2$-preparation balanced steady-state free precession ($T_2$bSSFP) have been used extensively (12–14). In recent years, there has been a growing interest in simultaneous $T_1$ and $T_2$ mapping techniques for the myocardium (15–22). Some of this work has been based on the magnetic resonance fingerprinting (MRF) paradigm (23–25). MRF involves devising a pulse sequence which yields different magnetization evolution patterns for tissues with different $T_1$ and $T_2$ (but it could also include other quantities), and find the best match between the measured signal and a simulated dictionary with known $T_1$ and $T_2$. Typically, such dictionary-based fingerprinting schemes rely on heavily undersampled spiral or other non-Cartesian trajectories (26–28), where a measurement on the magnetization evolution time-curve is obtained from each aliased spoke. However, dictionary-based parameter mapping may also be combined with a Cartesian trajectory, in which case fewer time-points along the magnetization evolution curve can be used for dictionary matching, only those corresponding to the acquisition of the k-space center (29). The advantage of using Cartesian as opposed to spiral sampling is lower susceptibility to $B_0$-inhomogeneity, no signal aliasing and simplified reconstruction. Therefore, a Cartesian sampling strategy for dictionary-based $T_1$ and $T_2$ mapping may be more readily incorporated into a clinical setting.

The purpose of this work was to implement and perform initial optimization on a new dictionary-based technique for myocardial $T_1$ and $T_2$ mapping using Cartesian sampling. The proposed simultaneous $T_1$ and $T_2$ mapping technique (termed Multimapping) was tested in phantoms, healthy subjects and patients with cardiovascular disease.

## Methods

### *Multimapping pulse sequence design*

The proposed Multimapping pulse sequence consists of 10 single-shot images, ECG-triggered to the mid-diastolic rest period, and incorporating inversion and $T_2$ preparation ($T_2$prep) pulses to introduce $T_1$ and $T_2$ sensitization, respectively. The image acquisition consists of a balanced steady-state free precession (bSSFP) sequence with Cartesian sampling. Imaging parameters for all experiments and simulations are: field-of-view = 320×320 mm, spatial resolution = 2×2 mm, slice thickness = 10 mm, nominal flip angle = 50°, bandwidth = 1076 Hz/pixel, TR = 2.3 ms, TE = 1.2 ms, SENSE factor = 2, linear profile order. Ten startup radiofrequency (RF) pulses are used with linearly increasing flip angles. An adiabatic inversion pulse is used with an assumed inversion efficiency of 0.94 for the in vivo experiments (30), corresponding to an inversion angle of 160°. For the $T_2$prep module, hard 90° tip down and up pulses are used with four adiabatic refocusing pulses in between.

The timings of the inversion and $T_2$prep pulses may be optimized to improve $T_1$ and $T_2$ sensitivity. Specifically, using more inversion pulses with different delays may increase sensitivity to different $T_1$, while using more $T_2$prep with different echo times may yield better $T_2$ sensitivity. To provide some preliminary optimization in this regard, Multimapping pulse sequences with different pre-pulse settings were implemented and evaluated, the details of which are provided in Supporting Information Section 1. The optimized Multimapping pulse sequence is illustrated in Figure 1, consisting of inversion pulses in cycles 1 and 5, and $T_2$prep modules in cycles 8, 9 and 10 with different echo times of 30, 50 and 70 ms, respectively.

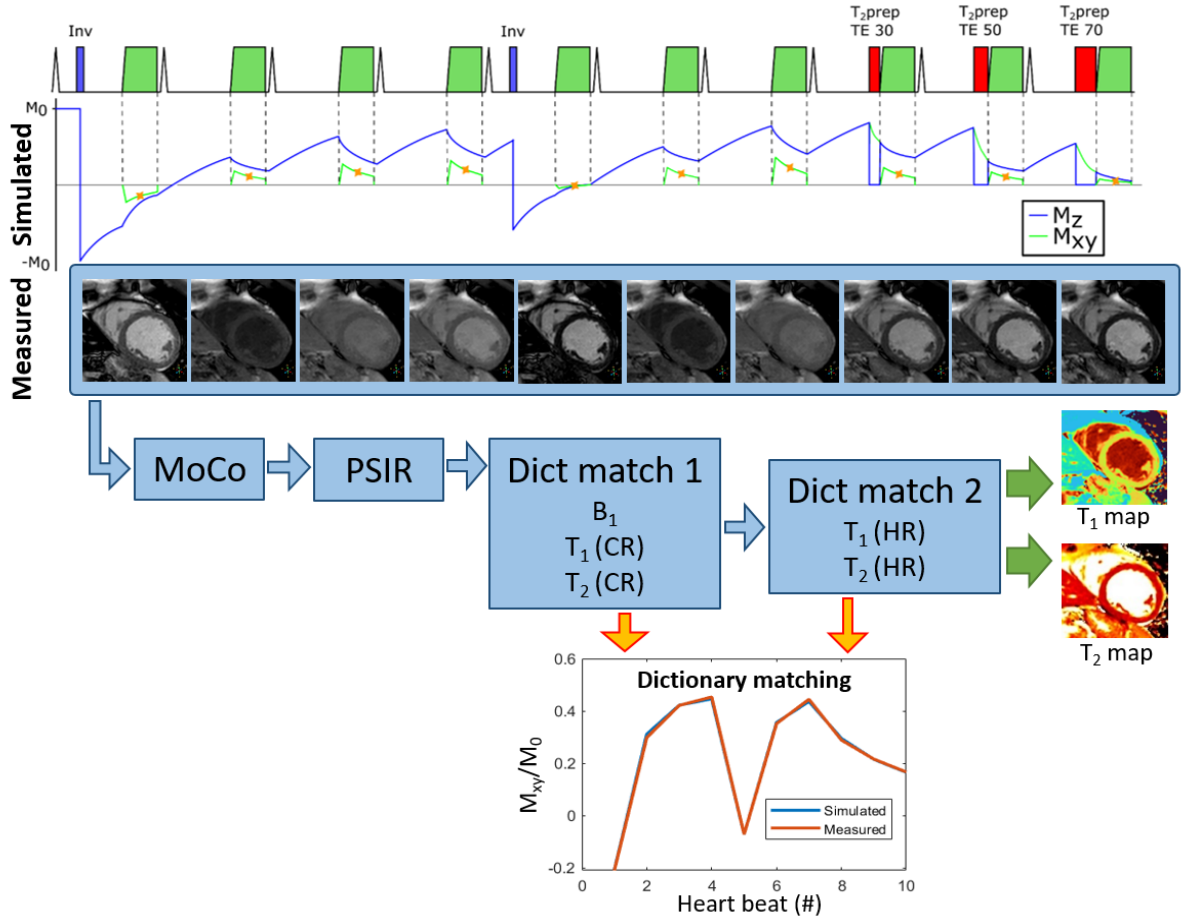

*Figure 1. Pulse sequence and post-processing steps for the proposed single-shot Cartesian Multimapping technique, acquired across 10 consecutive cardiac cycles, using inversion pulses and $T_2$prep modules with varying echo times to introduce $T_1$ and $T_2$ sensitization, respectively. After image reconstruction, motion correction (MoCo) is performed to spatially align the 10 source images, followed by phase sensitive inversion recovery (PSIR) correction to determine the polarity of the magnetization. Dictionary matching is then performed (Dict match 1) incorporating $B_1$, and coarsely resolved (CR) $T_1$ and $T_2$, to determine a global $B_1$ which is applied to the second dictionary matching step (Dict match 2) where highly resolved (HR) $T_1$ and $T_2$ maps are generated. For the dictionary matching, the acquired data at each pixel is matched to the dictionary entry which best matches the simulated sequence for a certain combination of $T_1$ and $T_2$ (and $B_1$ for the first dictionary matching step), to yield parametric maps. Because a Cartesian trajectory is used, only the simulated center of k-space is matched to the acquired signal.*

*Image post-processing, dictionary generation and feature extraction*

The Multimapping post-processing steps are summarized in Figure 1. The Multimapping dictionaries (denoted as [lower value: step size: upper value]) are generated by simulating the pulse sequence using the extended phase graph (EPG) framework for ranges of $T_1$, $T_2$ and $B_1$. The RR-intervals are recorded for each scan and, along with the patient-specific trigger delays, used to simulate a subject-specific dictionary. Unlike previous similar dictionary-based cardiac mapping techniques, which uses spiral or radial sampling to measure the transverse magnetization ($M_{xy}$) for each RF-pulse in the mid-diastolic acquisition window, here a Cartesian trajectory is used which results in a single $M_{xy}$ measurement per cardiac cycle. As a result, only the simulated $M_{xy}$ for the corresponding center of k-space acquisition in the cardiac cycle is used for the dictionary. To avoid the 'curse of dimensionality' (16) which results in extremely long dictionary generation times, the radiofrequency $B_1^+$ field ($B_1$) is first estimated using a relatively coarse grid for $T_1$ [500:100:1500] and $T_2$ [40:30:140] and a $B_1$ range of [0.5:0.05:1]. The estimated $B_1$ is calculated as the mean $B_1$ in a manually selected region of interest (ROI) in the ventricular septum and is used as input for the more highly resolved $T_1$ and $T_2$ dictionaries. These dictionaries which are only partially resolved for $T_1$ and $T_2$ are generated separately, where each dictionary contain a highly resolved and a coarsely resolved parameter, to reduce dictionary generation time. For the in vivo experiments, the partially resolved $T_1$ dictionary ($T_1$ Dict$_{PR}$) has a $T_1$ range of [200:1:2500] and $T_2$ range of [20:30:150], and the partially resolved $T_2$ dictionary ($T_2$ Dict$_{PR}$) has a $T_1$ range of [200:50:2500] and $T_2$ range of [1:1:150]. The $T_2$ and $T_1$ time-steps for the $T_1$ Dict$_{PR}$ and $T_2$ Dict$_{PR}$, respectively, were empirically determined to provide a reasonable trade-off between short processing time and low quantification error, using in vivo data from three healthy subjects. Details on this optimization are provided in Supporting Information Section 2. For the phantom experiments, the dictionary has a $T_1$ range of [1:10:2500] and $T_2$ range of [1:4:400]. MATLAB scripts for the dictionary generation and feature extraction can be downloaded from github, including example Multimapping source images from one healthy subject (https://github.com/Multimapping/Matlab_files).

Prior to dictionary matching, in vivo images are motion corrected using non-linear image registration to minimize respiratory or cardiac-induced motion. The motion correction procedure consists of a total variation regularization algorithm implementation with default parameters (31). A second processing step performs phase sensitive correction on the acquired images to determine the polarity of the signal (32). Dictionary matching is then performed using the dot product between the dictionary entries and the motion and polarity corrected images, similar to previous dictionary-based techniques (23).

### MRI experiments

All experiments were performed on a Philips 1.5T Ingenia scanner (Philips Healthcare, Best, The Netherlands) with a 28-channel cardiac coil. All subjects provided written informed consent and the study was approved by the local ethics committee. Imaging parameters for all experiments are outlined in the Multimapping pulse sequence section. Images were reconstructed on the scanner and transferred to an offline workstation with an Intel Core i7-8565U (1.80 GHz) processor and 16Gb RAM for post-processing using MATLAB (The MathWorks, Natick, MA). With this setup, the $B_1$ dictionary took approximately 10 seconds to generate, and $T_1$ Dict$_{PR}$ and $T_2$ Dict$_{PR}$ approximately 13 minutes in total.

### Phantom studies

The International Society for Magnetic Resonance in Medicine/National Institute of Standards and Technology (ISMRM/NIST) phantom was scanned to evaluate any heart-rate dependency (33,34). Multimapping scans were performed with different heart-rates, from 40 bpm to 120 bpm with 20 bpm increments, to determine any heart-rate dependency. Reference $T_1$ and $T_2$ values were obtained using inversion recovery spin echo (IRSE) for $T_1$, and multi-echo spin echo (MESE) for $T_2$. Imaging parameters for the reference scans were: 2×2 mm spatial resolution, 10 mm slice thickness, 8 inversion times from 50 to 2000 ms, 7000 ms TR for the IRSE experiment, and 8 echoes with TR/TE/$\Delta$TE = 7000/15/15 ms for the MESE experiment. Only vials with $T_1$ and $T_2$ values within physiologically relevant ranges (50 ms to 2200 ms for $T_1$ and 5 to 400 ms for $T_2$) were considered. To evaluate agreement between reference $T_1$ and $T_2$ values in the phantom and those measured with Multimapping, correlation coefficients ($R^2$) and normalized root mean square errors (NRMSE) were calculated.

Repeatability of the in vitro $T_1$ and $T_2$ quantification was assessed by performing a second Multimapping phantom experiment (Scan 2) two months after the experiments described in the previous paragraph. For this experiment, only a simulated heart rate of 60 bpm was performed and compared to the corresponding scan with the same simulated heart rate in the previous phantom experiments (Scan 1). Multimapping $T_1$ and $T_2$ values were compared between Scan 1 and Scan 2 by calculating $R^2$ and Bland-Altman analysis.

### In vivo experiments

In vivo experiments were performed in 16 healthy subjects (11 male, age: 30 ±5.3 years, heart rate: 65±6.4 bpm), one patient with dilated cardiomyopathy (47-year-old male, heart rate: 61 bpm) and two patients with acute myocarditis (one 49-year-old female, heart rate: 44 bpm; one 41-year-old female, heart rate: 58 bpm). A Multimapping slice was acquired in mid-ventricular short axis view in all subjects. Additionally, 5(3b)3 MOLLI was acquired for comparison of $T_1$ values, while $T_2$bSSFP with 4 different $T_2$ weighting (0, 23, 46, and 70 ms echo times, three pause cardiac cycles between each image) was acquired to compare $T_2$ values. Both MOLLI and $T_2$bSSFP were acquired in the same mid-ventricular short-axis location as the Multimap slice with identical imaging parameters, apart from the flip angle which was 35° for MOLLI and $T_2$bSSFP. MOLLI used an adiabatic inversion pulse, identical to the one used for Multimapping, while the $T_2$prep module for $T_2$bSSFP was also the same as the one for Multimapping. Vendor-provided online curve fitting algorithms were used for parameter estimation of both MOLLI and $T_2$bSSFP. To assess in vivo intra-scan repeatability, Multimapping, MOLLI and $T_2$bSSFP were repeated in six of the healthy subjects in the same scan session, waiting approximately 10 minutes between the acquisitions.

The Multimapping $B_1$ maps were validated in three healthy subjects in which $B_1$ mapping was performed using the single-shot dual refocusing echo mode (DREAM) technique (35,36), in addition to the Multimapping technique. Details of the $B_1$ mapping validation experiments are provided in Supporting Information Section 3.

To compare Multimapping parameter maps to MOLLI and $T_2$bSSFP, ROIs were manually drawn in the $T_1$ and $T_2$ maps according to the 6 mid-ventricular segments of the American Heart Association (AHA) model (37). Additionally, measurements were performed across the entire mid-ventricular slice. Differences in mean values and standard deviations (spatial variability) were analyzed using pair-wise t-tests with significance threshold of $P < 0.05$.

To investigate the influence of the $B_1$ correction, $T_1$ and $T_2$ maps generated using $B_1$ estimated from an ROI in the septum were compared to maps generated without correction ($B_1$=1). This comparison was performed by calculating the sum of squared differences (SSD) between the measured and the simulated signal with and without $B_1$ correction. $B_1$ was measured in the 6 AHA segments to evaluate homogeneity across the slice. The $B_1$ analysis was performed for all healthy subject datasets.

## Results

### *Phantom*

Multimapping phantom measurements (Figure 2) showed good agreement with reference values for both $T_1$ and $T_2$. There did not appear to be noticeable systematic errors in general for Multimapping $T_1$ or $T_2$ for the different heart-rates, although $T_2$ values above 120 ms were higher compared to reference values for heart-rates of 100 and 120 bpm. The $R^2$ between $T_1$ and $T_2$ estimated using Multimapping compared to reference values was above 0.99 for all simulated heart-rates. The $T_1$ NMRSE was largest for the heart-rate of 40 bpm at 4.2% with a range of 0.5% to 11.5%. The $T_2$ NMRSE was largest for the heart-rate of 120 bpm at 11.9% with a range of 2.4% to 32.9%. The $T_1$ and $T_2$ NMRSE and ranges for all simulated heart rates are summarized in Supporting Information Table S1.

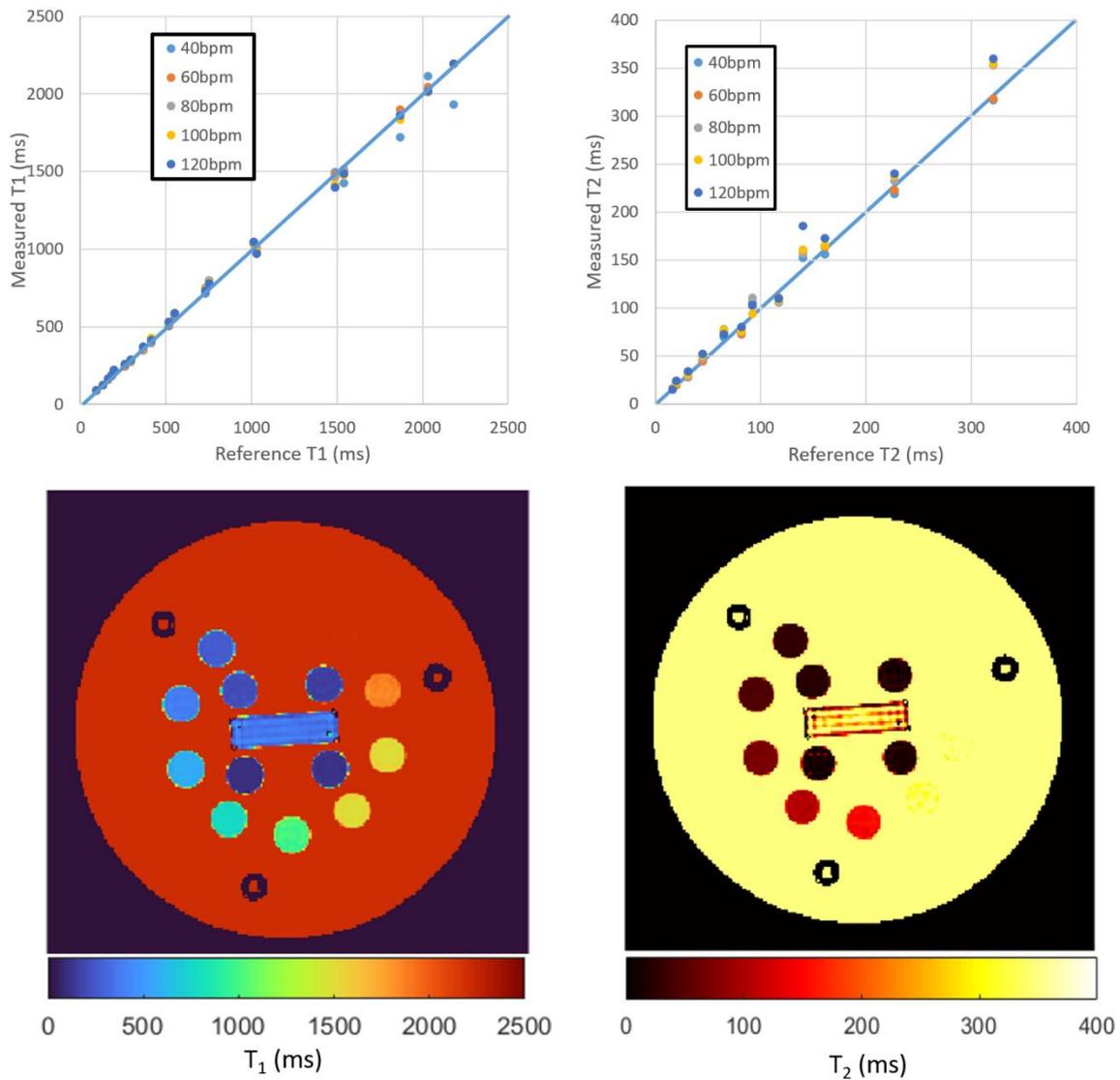

*Figure 2. $T_1$ and $T_2$ values measured with Multimapping in the ISMRM/NIST phantom and for different heart rates from 40 to 120 bpm, compared to reference values. A good agreement between reference and Multimapping $T_1$ was found without any discernable heart rate dependency. A good agreement for $T_2$ was also found for values less than approximately 100 ms. A slight over-estimation for higher $T_2$ was observed. The images on the bottom row show Multimapping $T_1$ and $T_2$ maps for the ISMRM/NIST phantom $T_2$ slice for a simulated heart rate of 60 bpm. Dictionary matching was only performed for the $T_1$ range of 50 to 2500 ms and a $T_2$ range of 5 to 400 ms which means that not all vials could be visualized.*

Repeatability assessment of Mutimapping $T_1$ and $T_2$ quantification yielded a $R^2$ of more than 0.99 with no significant bias ($T_1$ bias = -1.1 ms, $T_2$ bias = -1.2 ms). The correlation and Bland-Altman plot from the phantom repeatability experiments are shown in Supporting Information Figure S7.

### *In vivo*

Segment-wise $B_1$ measurements for all 16 healthy subjects are shown in Supporting Information Figure S8. The mean ± standard deviation (SD) relative $B_1$ across all segments and subjects was 0.60 ± 0.04 and the mean SD for each subject (intra-subject relative $B_1$ variability) was 0.02. An example demonstrating the usefulness of $B_1$ correction for Multimapping is shown in Figure 3. The figure shows in vivo $T_1$ and $T_2$ maps from one healthy subject incorporating $B_1$ measured in the septum, with lower difference between measured and simulated signal compared to no $B_1$ correction ($B_1$ = 1). The mean ± SD sum of squared distance of the simulated and measured magnetization (in $M_{xy}/M_0$) for the myocardium in the healthy subjects using $B_1$ correction was 0.0035 ± 0.0012, significantly smaller than for no $B_1$ correction (0.010 ± 0.0022; $p < 0.0001$). Without $B_1$ correction the mean $T_1$ across all 16 healthy subjects was 1128 ± 20 ms and $T_2$ was 57.5 ± 3.4, significantly higher than with $B_1$ correction ($T_1$ = 1114 ± 14 ms, $p < 0.01$; $T_2$ = 47.1 ± 1.3 ms, $p < 0.01$).

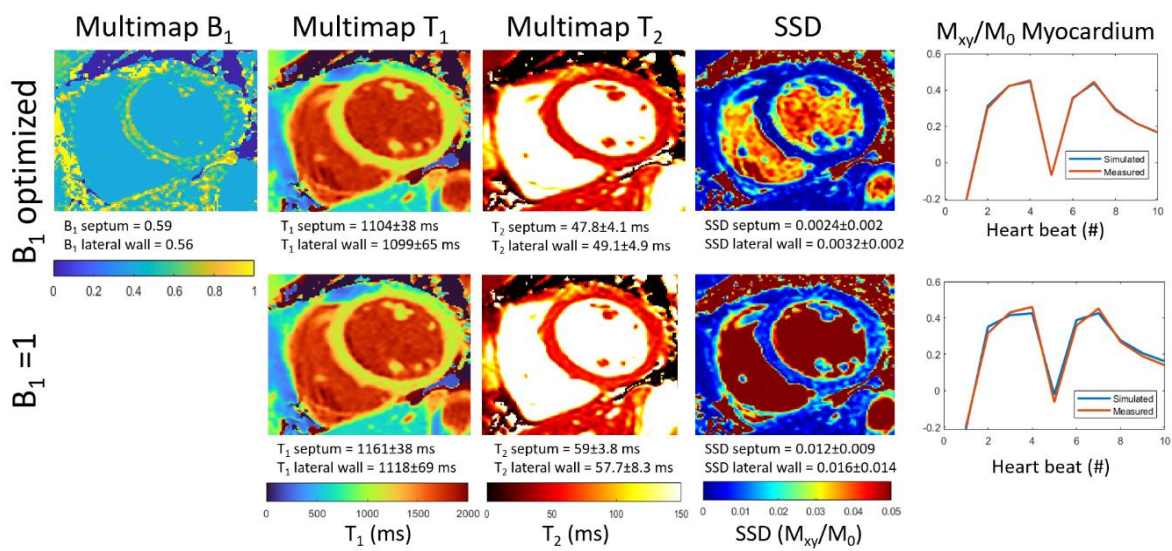

*Figure 3. Multimapping parameter maps from one healthy subject using $B_1$ optimized from a septal region-of-interest and applied to the $T_1$ and $T_2$ dictionaries (top row). The Multimapping $B_1$ map was first calculated using dictionary matching from coarsely sampled $T_1$ and $T_2$ in the range of native myocardium. For comparison, $T_1$ and $T_2$ maps were generated without $B_1$*

Multimapping $T_1$ and $T_2$ maps generated with and without respiratory motion correction are shown in Supporting Information Figure S9 from the healthy subject with the largest respiratory motion amplitude. Representative Multimapping $T_1$ and $T_2$ maps for two healthy subjects are shown in Figure 4, along with MOLLI and $T_2$bSSFP for comparison. Comparable image quality between the Multimapping images and reference techniques can be observed. Bullseye plots of the group mean $T_1$ and $T_2$ measurements for the healthy subjects are shown in Figure 5, including segmental and global measurements. Group-wise mean $T_1$ for the entire slice was significantly higher using Multimapping ($T_1 = 1114 \pm 14$ ms) compared to MOLLI ($T_1 = 991 \pm 26$ ms; $p < 0.01$). Furthermore, group-wise mean Multimapping $T_2$ ($47.1 \pm 1.3$ ms) was significantly lower compared to $T_2$bSSFP for the entire slice ($T_2 = 54.7 \pm 2.2$ ms; $p < 0.001$). Bland-Altman plots of myocardial $T_1$, blood $T_1$ and myocardial $T_2$ for the entire slice from Multimapping, MOLLI and $T_2$bSSFP are shown in Figure 6. Bullseye plots of the $T_1$ and $T_2$ mean spatial variability for Multimapping and reference techniques are shown in Figure 7. There was no difference in $T_1$ spatial variability between Multimapping ($63.5 \pm 12.8$ ms) and MOLLI ($66.1 \pm 15.4$ ms) for the entire slice ($p = 0.62$). However, $T_2$ spatial variability was significantly lower for Multimapping ($5.8 \pm 1.0$ ms) compared to $T_2$bSSFP ($8.4 \pm 2.0$ ms) for the entire slice ($p < 0.01$). A Multimapping source image, $T_1$ and $T_2$ map for one healthy subject are shown in Supporting Information Figure S10. Although $T_1$ and $T_2$ are homogeneous across the myocardium, quantification differences, particularly for $T_2$, can be observed for peripheral subcutaneous fat.

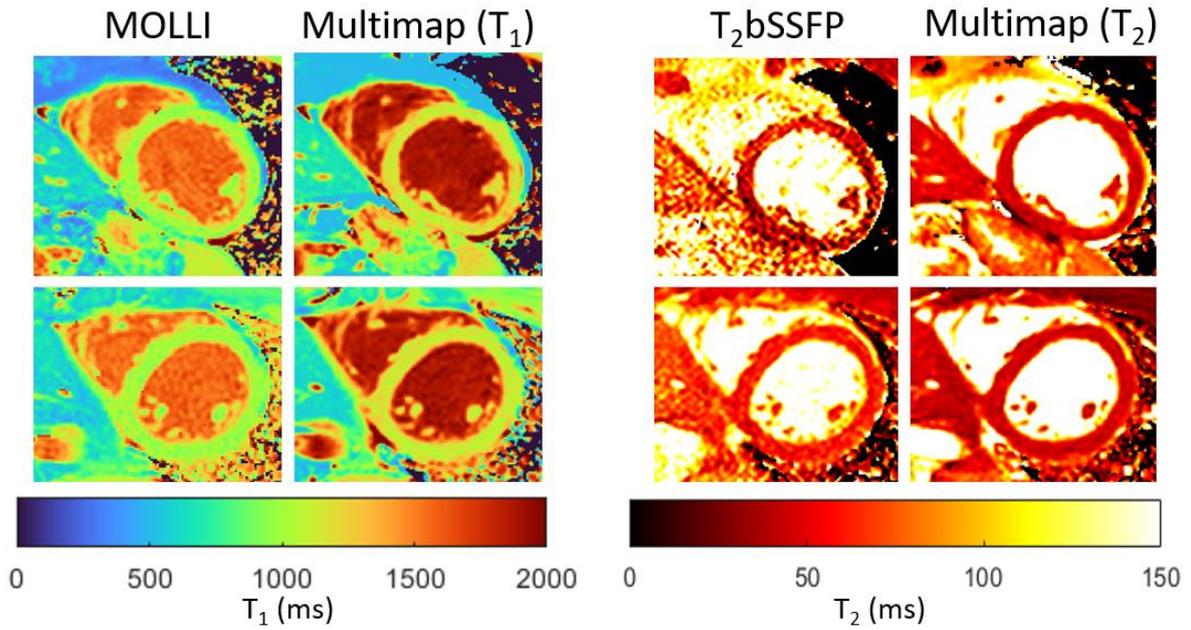

*Figure 4. $T_1$ and $T_2$ Multimaps and reference maps for two healthy subjects. The image quality is similar for the Multimaps and reference techniques, with a slightly increased signal-to-noise ratio for the $T_2$ Multimap compared to the reference technique ($T_2$bSSFP).*

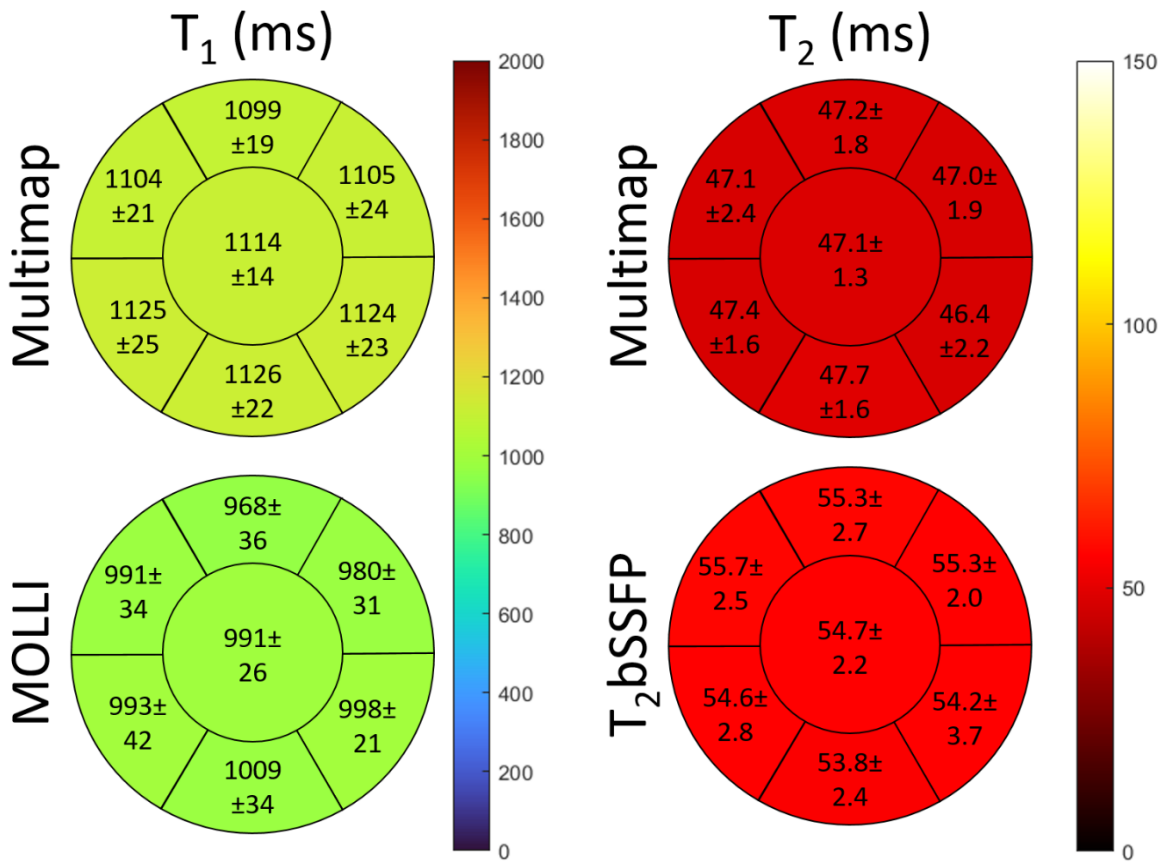

*Figure 5. Bullseye plots of the group-wise mean ± standard deviation for the mean $T_1$ and $T_2$ measured in the mid ventricular slice for the 6 AHA segments and for the entire slice (in the center of each bullseye). Multimapping mean $T_1$ was higher compared to MOLLI $T_1$, but $T_2$ was lower than $T_2bSSFP$.*

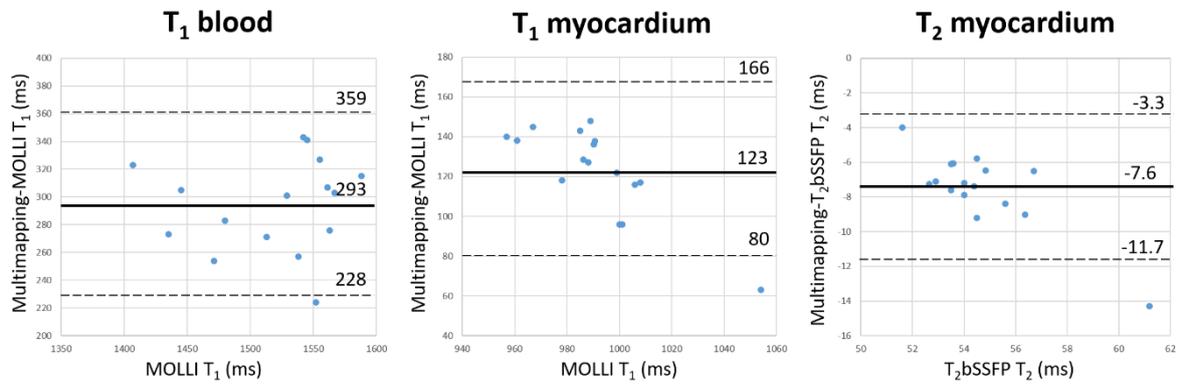

*Figure 6. Bland-Altman plots comparing Multimapping to the in vivo reference techniques. The thick line indicates the bias and thin dashed lines the 95% limits of agreement.*

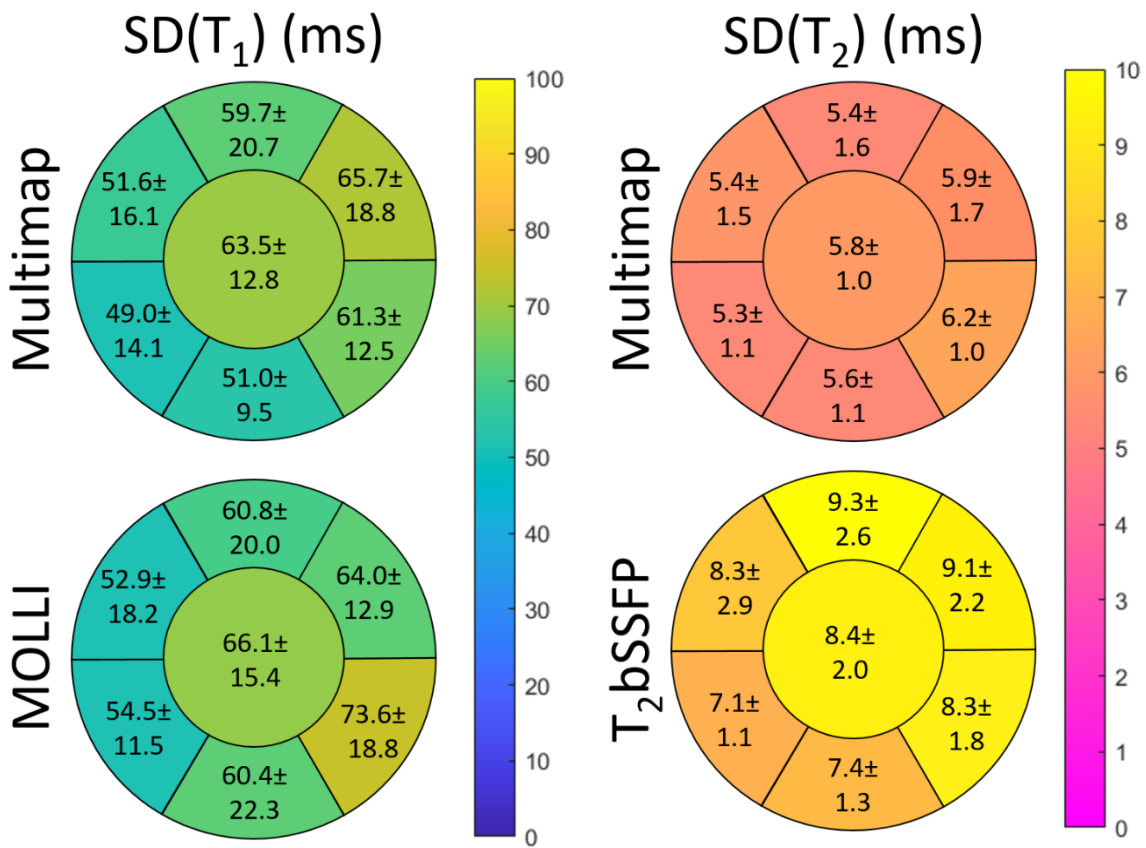

*Figure 7. Bullseye plots of the group-wise mean ± standard deviation for the spatial variability (as measured with the standard deviation) of $T_1$ and $T_2$ measured in the mid ventricular slice for the 6 AHA segments and for the entire slice (in the center of each bullseye). Multimapping spatial variability for $T_1$ was similar as MOLLI, but $T_2$ spatial variability was lower than that of $T_2bSSFP$.*

Bland-Altman plots comparing the repeated Multimapping, MOLLI and $T_2bSSFP$ scans in six healthy subjects, for the six AHA segments (and left ventricular blood pool for $T_1$), are shown in Figure 8. All techniques yielded biases less than 1 ms between the repetitions. However, the 95% limits of agreement were smaller for Multimapping $T_1$ (-24.7 to 26.3 ms) compared to MOLLI (-34 to 32.3 ms), and Multimapping $T_2$ (-1.6 to 1.6 ms) compared to $T_2bSSFP$ (-2.9 to 2.4 ms).

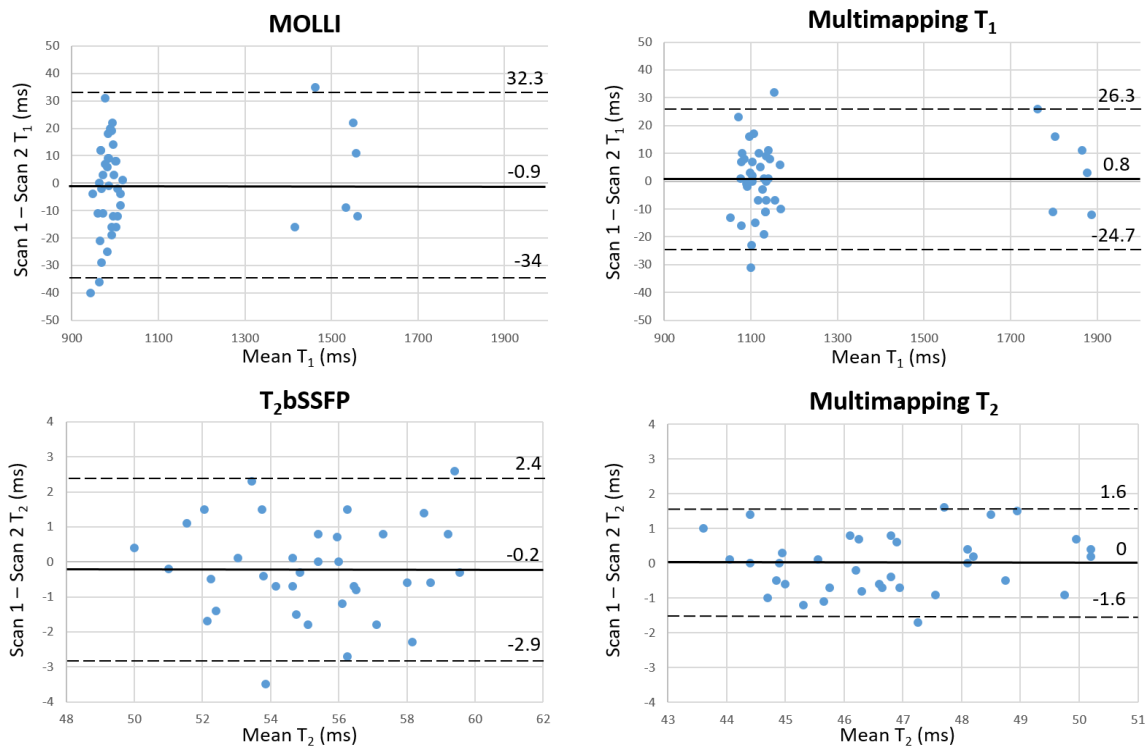

*Figure 8. Bland-Altman plots of repeated MOLLI, $T_2bSSFP$ and Multimapping scans from six healthy subjects with $T_1$ and $T_2$ measured for each of the 6 mid-ventricular AHA segments and left ventricular blood pool for $T_1$.*

$T_1$ maps, $T_2$ maps and late gadolinium enhancement (LGE) images from the patient with dilated cardiomyopathy are shown in Figure 9. Septal $T_1$ and $T_2$ from the Multimap (which overlaps with a region of LGE) was increased in this patient compared to the values in the cohort of healthy subjects. Increased $T_1$ but not $T_2$ was measured with the reference techniques, although the $T_1$ increase was more modest compared to the difference measured with Multimapping. Multimaps, reference maps and LGE images for the 49-year-old patient with acute myocarditis (Figure 10) reveal an area of acute inflammation in the anterior segment. Multmapping $T_1$ in this area was 1373 ms, compared to 1111 ms in the inferolateral segment without enhancement. Corresponding MOLLI $T_1$ was 1267 ms in the anterior segment, and 1004 ms in the inferolateral segment. Multimapping $T_2$ was 72.9 ms and 46.4 ms in the anterior and inferolateral segments, respectively. Corresponding $T_2$bSSFP $T_2$ was 82.5 ms and 57.3 ms, respectively. Multimaps, LGE and short tau inversion recovery (STIR) images for the 41-year-old patient with acute myocarditis are show in Supporting Information Figure S11. In this patient the $T_1$ and $T_2$ values were significantly increased globally. However, due to the diffuse and global nature of the disease in this patient, myocardial contrast agent uptake was relatively difficult to appreciate in the $T_1$-weighted LGE images, while the $T_2$-weighted STIR images also yielded an apparent constant signal which was only borderline pathological (signal intensity ratio of myocardium versus skeletal muscle = 2.0).

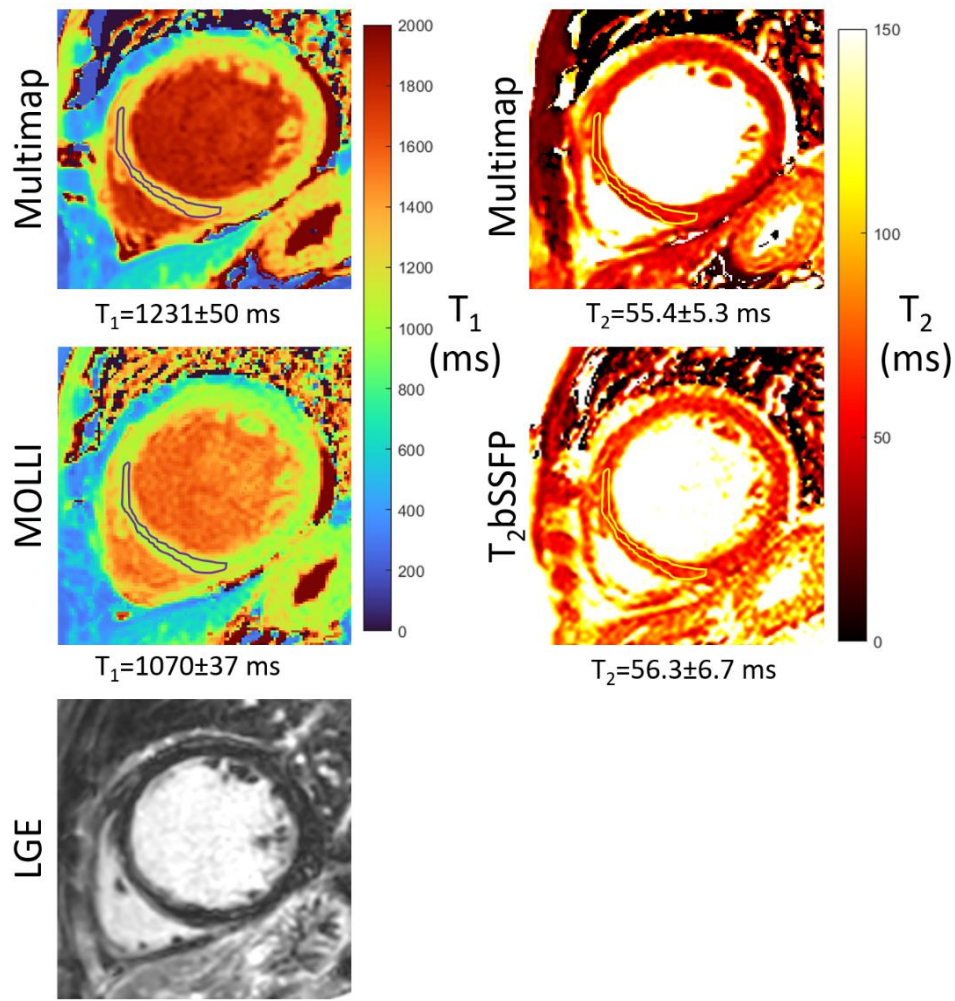

*Figure 9. Multimap, reference maps and a late gadolinium enhancement (LGE) image for the patient with dilated cardiomyopathy. $T_1$ and $T_2$ measured with Multimapping in the septal area with LGE is increased compared to the ranges measured in the healthy subjects for $T_1$ and $T_2$. For the reference techniques, a modest $T_1$ increase is measured with MOLLI, but $T_2$ measured with $T_2bSSFP$ is within the normal range.*

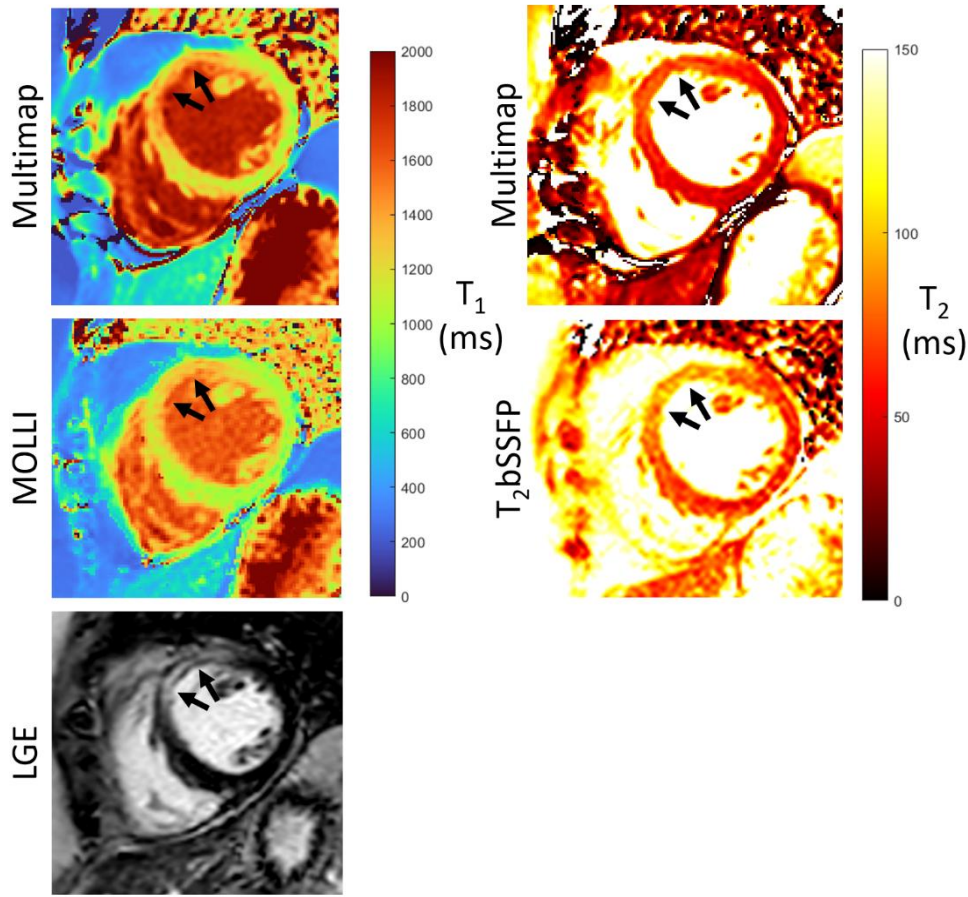

*Figure 10. Multimaps and reference $T_1$ and $T_2$ maps for the 49-year-old patient with acute myocarditis. An area of acute inflammation is seen in the anterior segment (arrows) in the $T_1$ and $T_2$ maps and LGE image.*

## Discussion

In this work, the single-shot Cartesian Multimapping technique was proposed for simultaneous myocardial $T_1$ and $T_2$ mapping using dictionary-matching. Multimapping yields generally accurate $T_1$ and $T_2$ times independently of the heart rate, as demonstrated in the phantom study. Studies in healthy subjects showed consistently different $T_1$ and $T_2$ values between Multimapping and reference techniques, although particularly for $T_1$ this is to be expected due to the known bias of MOLLI which underestimates $T_1$ (11,38).

Inversion pulses and $T_2$prep modules are frequently used for myocardial mapping techniques to provide $T_1$ and $T_2$ sensitization, respectively. The configuration of these pulses may be modified and optimized to increase accuracy and precision (30,39). A number of Multimapping embodiments were investigated in simulations and in vivo, all restricted to 10 cardiac cycle acquisitions, but with different inversion and $T_2$prep settings. There was a good agreement between simulation and in vivo results, showing that the highly $T_1$ sensitized approach fared better for $T_1$ quantification, while $T_2$ quantification suffered. Conversely, sensitizing primarily for $T_2$ yielded good $T_2$ quantification at the expense of $T_1$. The 'balanced' approach with two inversion pulses and three $T_2$prep modules appears to provide a good tradeoff between $T_1$ and $T_2$ quantification error. Nevertheless, this small optimization study was limited in scope and extent. Further experiments, using a more exhaustive range of Multimapping embodiments should be considered, which could also include using different or varying (between images) flip angles.

An assumption of the proposed Multimapping approach is that the $B_1$ field is homogeneous and can be represented by an ROI in the septum. Evaluation of $B_1$ across the healthy subjects scanned in this study and using a reference $B_1$ mapping technique suggests this is a valid assumption, with little variation across the left ventricle. However, the inter-subject $B_1$ variability was significant and approximately twice as large as the intra-subject variability, which justifies the use of a subject-specific $B_1$ determination rather than an empirically determined $B_1$ for all subjects. Incorporating subject-specific $B_1$ in the proposed fashion, as a pre-processing step, reduces the computational complexity of the dictionary matching, keeping the generation times low while avoiding overfitting the data and reducing precision. However, further studies are required to evaluate if this approach is valid in a more varied cohort of subjects, across vendor platforms, and importantly at different field strengths. Field inhomogeneity is typically more pronounced at 3T and an alternative strategy may be needed to ensure accurate quantification. Furthermore, the need for manual interaction to define the septal $B_1$ measurement is an obvious limitation of the technique, but one which could be solved using automation (40). Signal variations caused by $B_0$ inhomogeneities were also not considered in this work. The bSSFP technique used for Multimapping image acquisition is sensitive to $B_0$ field inhomogeneities, which may affect quantification. This was seen particularly in the image periphery, where $B_0$ is less homogeneous and $T_2$ quantification of subcutaneous fat yielded differences of approximately 40% (Supporting Information Figure S10). However, myocardial $T_1$ and $T_2$ did not suffer noticeably from such spatial variability, as very similar $T_1$ and $T_2$ was measured across the myocardium in the healthy subjects. Again, $B_0$ inhomogeneity is likely a greater source of errors at 3T, and strategies to mitigate against this should be considered such as including $B_0$ in the dictionary matching or using a spoiled gradient echo readout.

The emphasis of this work has been to develop a dictionary-based cardiac parameter mapping technique, similar to previously proposed cardiac MRF, but without the requirement for complex and time-consuming post-processing. This may potentially lower the barrier for clinical adoption of simultaneous multi-parametric mapping. Previous efforts in cardiac parameter mapping (particularly $T_1$ mapping) has resulted in standardized imaging parameters with the aim of producing comparable quantification across platforms, and this could be readily translated to Multimapping which uses a nearly identical Cartesian bSSFP image acquisition scheme. A long-standing challenge in conventional $T_1$ and $T_2$ mapping, where mapping is performed as separate scans, is the relatively large inter-subject variability of the quantification. Although some of this variability may be attributed to factors such as age, sex or even hydration status (41,42), $T_1$ mapping may be confounded by $T_2$ effects, particularly if bSSFP readouts are used, and vice-versa $T_2$ mapping can be confounded by $T_1$ differences in tissue (43). An advantage of simultaneous $T_1$ and $T_2$ mapping such as Multimapping is that these confounders are minimized as they are both included in the signal model. This may explain why the inter-subject variability of both $T_1$ and $T_2$ for Multimapping were lower than for the reference in vivo techniques.

Although the mean Multimapping $T_1$ of 1114 ms was higher than for MOLLI, it is still just under 100 ms shorter than reported values in what is considered accurate techniques such as SASHA or SAPPHIRE at 1.5T (44,45). Correction for sources of systematic errors such as slice profile correction may yield $T_1$ values in a similar range, and will be the focus of future work.

Nevertheless, the in vivo precision of Multimapping for quantifying myocardial $T_1$, as indicated by the spatial variability, was comparable to MOLLI, which is considered the most precise of the widely used $T_1$ mapping methods (11). There are several factors which influence the precision of a cardiac $T_1$ mapping method, including which kind of pre-pulse is used (inversion or saturation) and how many source images are used for the curve fitting or dictionary matching procedure (11,46–48). In both these aspects MOLLI and Multimapping are very similar, which could explain the near equivalence in $T_1$ precision. For $T_2$, there was a significant difference in mean values between Multimapping and the reference technique of -7.6 ms. There is less consensus regarding the true $T_2$ for healthy myocardium at 1.5T, which ranges from approximately 46 to 58 ms, depending on the technique or vendor used (49–52). However, for previous comparable cardiac MRF techniques, $T_2$ values in the range of 41 to 45 ms have been reporter (28,53–55), which is lower than the $T_2$ observed with Multimapping of 47.1 ms. Compared to $T_2$bSSFP, Multimapping produced $T_2$ maps with significantly less spatial variability. Similar as for $T_1$ mapping, $T_2$ precision for $T_2$bSSFP is related to the number of source images used which is typically limited by the need for long waiting times to allow for near complete $T_1$ recovery between image acquisitions (56). The lower spatial variability of Multimapping $T_2$ may be explained by the use of more images for the mapping procedure. While $T_2$bSSFP only used four source images for the fitting, 10 images were used for Multimapping. Even if only three cardiac cycles were preceded by $T_2$-preparation to increase $T_2$ sensitization in the Multimapping pulse sequence, the bSSFP readout itself provide $T_2/T_1$ contrast which is exploited by the dictionary-based Multimapping technique but not with $T_2$bSSFP (57).

The ability of Multimapping to detect pathological changes in $T_1$ and $T_2$ was demonstrated in three patients, one with dilated cardiomyopathy and two with acute myocarditis. In all cases a significant increase in $T_1$ and $T_2$, well above the healthy ranges, was observed. This finding is consistent with previous studies using mapping techniques in these patient cohorts (58–60). Although a similar increase was also found using the reference techniques in both patients with acute myocarditis, $T_2$bSSFP did not find a significant $T_2$ increase in the patient with dilated cardiomyopathy compared to the healthy subjects. Further studies are ongoing in patients with various cardiovascular diseases to compare Multimapping parameter values to clinical reference techniques and values obtained in healthy controls.

Long dictionary generation times is a significant impediment to the integration of dictionary-based techniques into the clinical workflow (25). To limit dictionary generation times while providing $T_1$ and $T_2$ maps with high parameter granularity, an approach using partially resolved dictionaries was proposed in this study. Comparison with a reference dictionary showed that this can be a practical solution to this problem, as $T_1$ and $T_2$ maps could be generated on a standard computer without optimized Matlab code within minutes, with excellent agreement for $T_2$ and good agreement for $T_1$ in the myocardium. Note, that this evaluation specifically considered native $T_1$ and $T_2$ for myocardium and may not be directly extended to tissues with other $T_1$ and $T_2$ values.

The proposed Multimapping technique shares similarities, but also has important differences compared to recent cardiac MRF techniques (27,28). Like cardiac MRF, inversion pulses and $T_2$-preparation are combined in the same pulse sequence and are the primary sources of $T_1$ and $T_2$ sensitization, respectively. However, Multimapping does not suffer from the signal aliasing that is intrinsic to non-Cartesian MRF, which has some important implications. The number of data points that can be used to match measurements to simulations for each dictionary entry is smaller for the proposed Cartesian approach by nearly two orders of magnitude (10 measurements for Multimapping versus approximately 1000 for conventional cardiac MRF). Although the number of acquired images could be increased to mitigate against overfitting, this would either increase the breath-hold duration or require the use of free-breathing acquisition strategies (15,53). An advantage of avoiding signal aliasing in the measurements, particularly for cardiothoracic and abdominal application, is that tissue outside the ROI such as subcutaneous fat does not degrade the measurements. Solutions to this problem for cardiac MRF have relied on non-Cartesian multi-echo acquisition strategies which lowers the acquisition efficiency and requires more complicated reconstruction algorithms (26,28). A Cartesian dictionary-based technique similar to Multimapping, proposed by Kvernby et al. (29), uses a simpler Bloch equation formulation which is limited to spoiled gradient echo readouts and has a signal-to-noise ratio penalty compared to the bSSFP readout used in this study. Another comparable Cartesian dictionary-based technique using EPG simulations to enable bSSFP readouts was recently proposed for $T_2$ mapping of the prostate (61), which uses a segmented k-space acquisition but demonstrates the feasibility of this approach. Finally, a dictionary-based Cartesian approach for myocardial $T_1$ and $T_2$ mapping was recently proposed by Milotta et al. using EPG simulations to generate the dictionaries (39). However, that study used a spoiled gradient echo, segmented k-space trajectory acquired during free breathing which combines data across multiple cardiac cycles which are subject to variability and susceptible to cardiac and respiratory motion artifacts. Conversely, the proposed approach uses a bSSFP single-shot trajectory acquired in a breath-hold, where the inversion delays and timings are tailored and specific to each image to account for heart rate variability or arrhythmia.

This study has several limitations. Only a small number of healthy subjects and a few patients were included in this study to demonstrate proof-of-concept. Larger studies in healthy subjects and patients with cardiovascular disease are required to establish quantification ranges to discriminate between healthy and diseased myocardium. A very limited in vivo intra-scan repeatability evaluation was performed in this study and a more extensive evaluation, including reproducibility and inter-scan repeatability, is required to comprehensive assess its clinical potential. Although the reference techniques included in this study, MOLLI and $T_2$bSSFP, may be considered the most clinically used and relevant, there are alternative mapping techniques which provide improvements in accuracy or precision which may be more appropriate reference techniques to assess these important quantification error metrics (44,49,62–64). The phantom used for in vitro experiments covers a wide range of $T_1$ and $T_2$, including those of pre- and post-contrast myocardium and blood. However, the phantom does not include the specific $T_1/T_2$ (~1200/50 ms) combinations found in the myocardium, which may have implication for assessing the accuracy of myocardial $T_1$ and $T_2$ quantification. In vitro studies using phantoms with similar $T_1/T_2$ combination as the myocardium will be performed in the future (65). Correction for confounders such as slice profile, inversion imperfections or $B_1$ were not considered in this work. Previous studies using MRF have demonstrated how this may be incorporated into the dictionary to improve accuracy at the expense of increasing dictionary generation times (27,28,66,67). Although this could improve the accuracy of the technique, it may also reduce precision. This consideration, in addition to the desire to limit the dictionary generation time to a short duration, was the rationale for only including $T_1$ and $T_2$ in the main dictionaries. Further work will explore if accuracy can be increased without compromising on precision or post-processing durations, for example by using machine learning (68).

## Conclusion

Multimapping allows simultaneous $T_1$ and $T_2$ mapping of the myocardium with a Cartesian trajectory, demonstrating good agreement with reference techniques in vitro and promising in vivo image quality and parameter quantification results. The post-processing time could be significantly reduced by generating partially resolved $T_1$ and $T_2$ dictionaries, and the Multimapping approach for acquisition and post-processing is well-suited for integration into clinical routine. Further studies using Mugltimapping in healthy subjects and patients with cardiovascular disease are warranted.

## Data availability statement

To support the findings of the manuscript, all MATLAB source code for generating the Multimapping dictionaries and feature extraction are available at https://github.com/Multimapping/Matlab_files, including example Multimapping source images for one healthy subject.

# Supporting Information 1: Magnetization preparation pulse optimization

Four embodiments of the Multimapping pulse sequence were implemented and evaluated to investigate the effects of different magnetization preparation configurations. MM NoPrep does not contain any pre-pulses which should yield poor $T_1$ and $T_2$ sensitivity. MM 3I-2$T_2$p, which is sensitized towards $T_1$, consists of three inversion pulses spaced three cardiac cycles apart, with variable inversion times, followed by two cardiac cycles with $T_2$prep echo times of 30 and 70 ms, respectively. The more $T_2$-sensitized MM 1I-5$T_2$p consists of a single inversion pulse, in the second cardiac cycle, in order to sample the baseline magnetization in the first cycle. The five last cycles are acquired with different $T_2$prep echo time, from 30 to 110 ms with 20 ms increments. MM 2I-3$T_2$p consists of two inversion pulses, spaced four cycles apart and with fixed delays. The last three cycles are acquired with $T_2$prep with different echo times of 30, 50 and 70 ms, respectively.

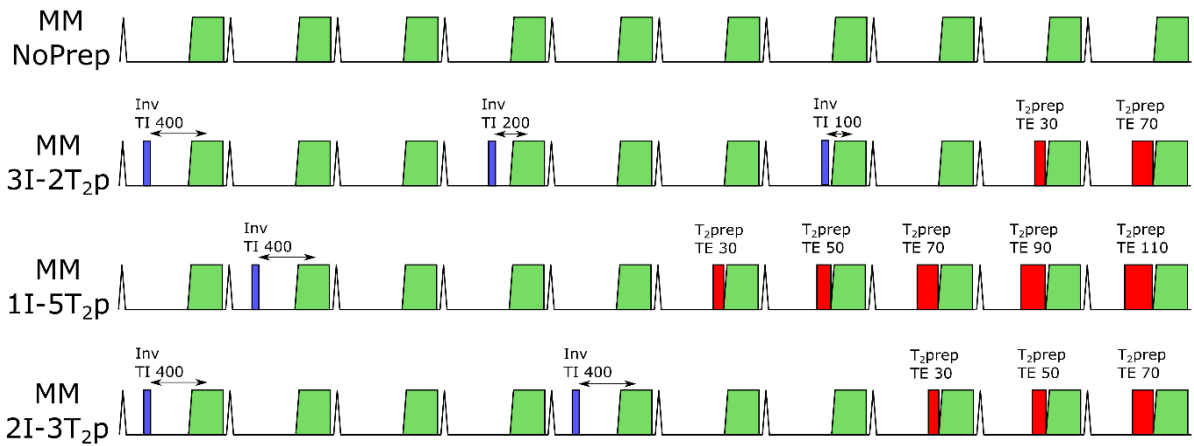

*Figure S1. Pulse sequence diagrams of different versions of the Multimapping techniques with variations of the magnetization preparation pulses (pre-pulses) to manipulate $T_1$ and $T_2$ sensitization. MM NoPrep does not contain any pre-pulses which should yield poor $T_1$ and $T_2$ sensitivity. MM 3I-2$T_2$p is particularly sensitized to $T_1$ and contain three inversion pulses at different inversion times and two $T_2$ preparation modules with echo times of 30 and 70 ms. MM 1I-5$T_2$p is sensitized to $T_2$ and contains a single inversion pulse and five $T_2$ preparation modules, with echo times ranging from 30 to 110 ms with 20 ms increments. Finally, MM 2I-3$T_2$p aims to balance $T_1$ and $T_2$ sensitivity and consist of two inversion pulses and three $T_2$ preparation modules with echo times of 30, 50 and 70 ms.*

Simulations were performed to compare the four candidate Multimapping pulse sequences using a numerical phantom, with $T_1$ and $T_2$ values of myocardium ($T_1$=1200 ms, $T_2$=50 ms) and blood ($T_1$=1800 ms, $T_2$=250 ms) modified for a field strength of 1.5T. The heart rate was set to 80 bpm. The root mean square error was calculated for the myocardium and used to compare the simulated scans. Images from the numerical phantom, simulating the four candidate Multimapping pulse sequences, are shown in Figure S2. A high RMSE was found for MM NoPrep of 137 ms for $T_1$ and 25 ms for $T_2$. While a low $T_1$ RMSE of 0.7 ms was measured for MM 3I-2$T_2$p, the $T_2$ RMSE was higher at 6.2 ms. Conversely, MM 1I-5$T_2$p yielded RMSE of 42 ms and 0.2 ms for $T_1$ and $T_2$, respectively. MM 2I-3$T_2$p resulted in a $T_1$ and $T_2$ RMSE of 2.3 ms and 0.6 ms, respectively.

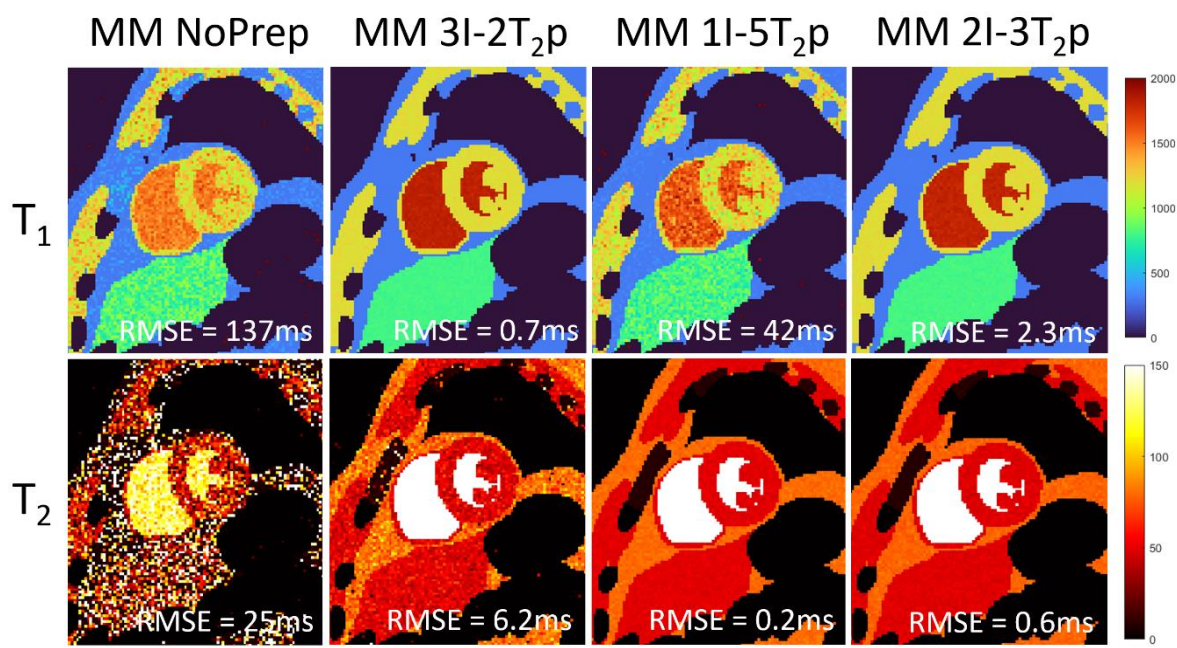

*Figure S2. Simulations of the Multimapping pulse sequences illustrated in Figure S1 for a numerical phantom of the heart. Differences in configuration of inversion and $T_2$ preparation pulses lead to different root mean square errors (RMSE) for $T_1$ and $T_2$ of myocardium. A comparably low RMSE for both $T_1$ and $T_2$ is achieved using Multimapping MM 2I-3$T_2$p.*

Three Multimapping pulse sequences (MM 3I-2$T_2$p, MM 1I-5$T_2$p, MM 2I-3$T_2$p) were performed in two healthy subjects to evaluate in vivo the influence of the different pre-pulse settings. MM NoPrep was not performed in vivo as it was considered unlikely to provide reasonable parameter maps. Multimapping $T_1$ and $T_2$ maps from one healthy subject comparing MM 3I-2$T_2$p, 1I-5$T_2$p and 2I-3$T_2$p are shown in Figure S3. A similar mean myocardial $T_1$ was measured for both MM 3I-2$T_2$p (subject 1=1099 ms, subject 2 = 1109 ms) and MM 2I-3$T_2$p (1097 ms, 1099 ms), but a higher myocardial $T_1$ was found for MM 1I-5$T_2$p (1158 ms, 1174 ms). In both subjects, a slightly lower myocardial $T_1$ spatial variability was found for MM 3I-2$T_2$p (50 ms, 41 ms) compared to MM 2I-3$T_2$p (65 ms, 43 ms), but both were much lower than MM 1I-5$T_2$p (82 ms, 65 ms). For mean $T_2$, there was no noticeable trend for higher or lower $T_2$ for any of the sequences. However, T₂ spatial variability was highest for MM 3I-2T₂p (6.7 ms, 5.6 ms), and lowest for MM 1I-5T₂p (4.9 ms, 4.4 ms), with MM 2I-3T₂p in between (5.3 ms, 5.1 ms).

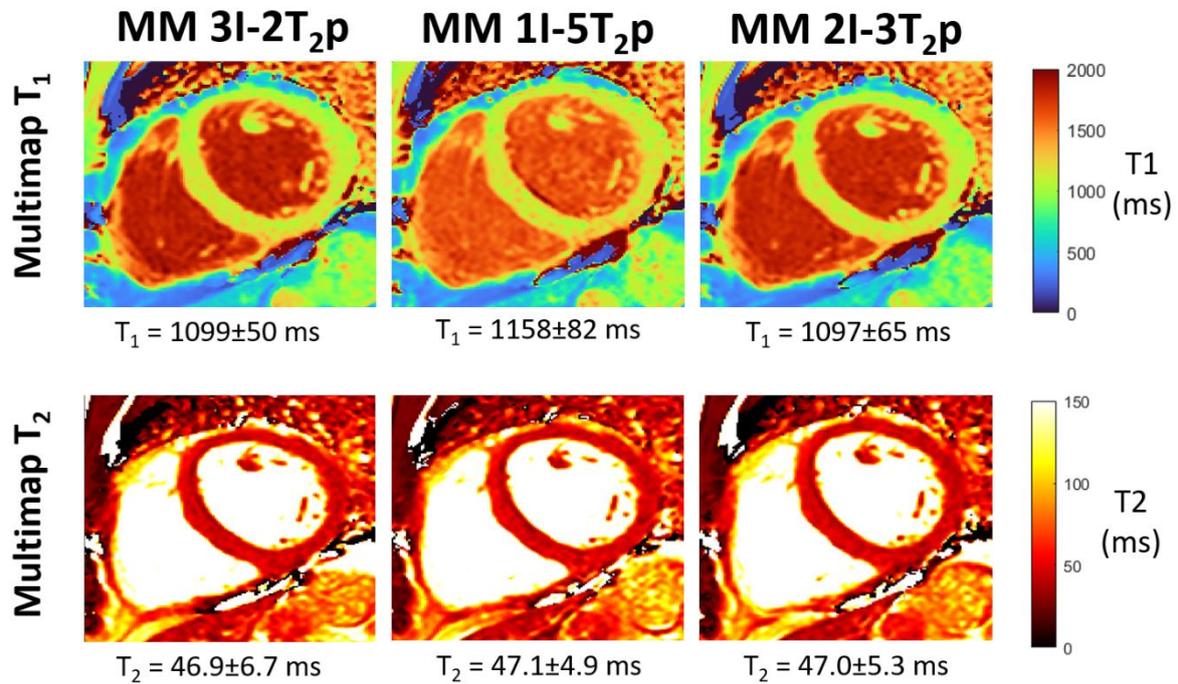

*Figure S3. $T_1$ and $T_2$ Multimaps generated using the 3I-2T₂p, 1I-5T₂p and 2I-3T₂p schemes in one healthy subject. 3I-2T₂p and 2I-3T₂p generate similar $T_1$ maps, although the latter scheme has a slightly higher spatial variability as indicated by the standard deviation. In comparison, 1I-5T₂p yields higher myocardial mean $T_1$ and spatial variability. Mean $T_2$ was approximately the same for all schemes, while $T_2$ spatial variability was highest for 3I-2T₂p and lowest for 1I-5T₂p.*

## Supporting Information 2: Partial dictionary error versus processing time

To minimize the processing time for the scan specific $T_1$ and $T_2$ dictionaries, partially resolved dictionaries were generated which had a highly resolved and a coarsely resolved $T_1$ or $T_2$ component. Using a highly resolved dictionary for both $T_1$ and $T_2$ that were simulated for three healthy subjects as references, normalized root mean square errors (NMRSE) for partially resolved dictionaries with different $\Delta T_1$ and $\Delta T_2$ time steps were calculated. For the partially resolved $T_1$ dictionaries, $\Delta T_2$ steps from 10 to 50 ms with 10 ms increments were evaluated. For the partially resolved $T_2$ dictionaries, $\Delta T_1$ steps of 10, 25, 50, 100 and 200 ms were evaluated. NMRSE was calculated for each pixel in the myocardium of the three $T_1$ and $T_2$ datasets, considering the difference between maps from the reference dictionaries and the partially resolved dictionaries and normalized by the expected $T_1$ (1100 ms) and $T_2$ (48 ms). The mean NMRSE was averaged across the myocardium for each subject and the final NMRSE for each time-step was the average across the three subjects. The highly resolved $T_1$ and $T_2$ reference dictionary had $T_1$ and $T_2$ ranges centered around healthy myocardium (900 to 1500 ms for $T_1$ and 20 to 80 ms for $T_2$) to minimize computational time and memory requirements. All processing was implemented in MATLAB (The MathWorks, Natick, MA) and performed on an Intel Core i7-8565U (1.80 GHz) processor with 16Gb RAM. The dictionary generation time was recorded for all partially resolved dictionaries. For both partially resolved $T_1$ and $T_2$ dictionaries a NMRSE less than 0.5% for either parameter was considered acceptable. As shown in Figure S1, this corresponds to $\Delta T_2$=30 ms for the $T_1$ Dict$_{PR}$ and $\Delta T_1$=50 ms for the $T_2$ Dict$_{PR}$. With these settings the $T_1$ Dict$_{PR}$ takes approximately 460 seconds and the $T_2$ Dict$_{PR}$ approximately 310 seconds to generate.

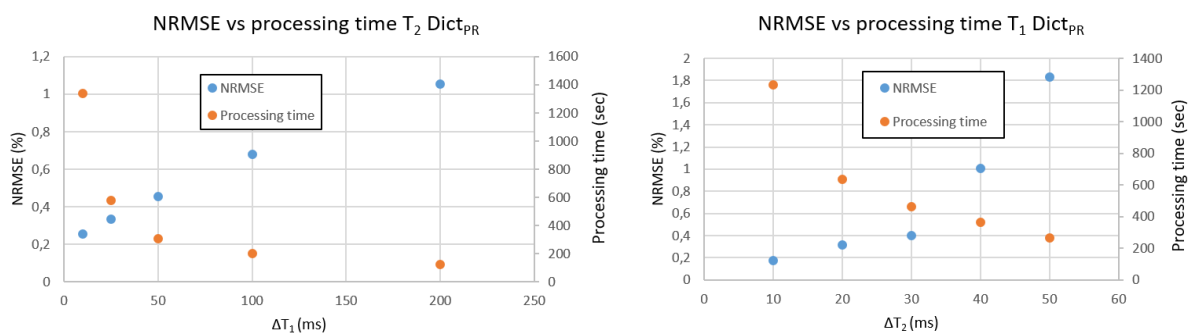

*Figure S4. Normalized root mean square error (NRMSE) versus processing time for the $T_2$ partial dictionary (left) and NRMSE versus processing time for the $T_1$ partial dictionary (right).*

## Supporting Information 3: $B_1$ mapping validation

To validate the Multimapping $B_1$ maps, additional reference $B_1$ maps were acquired in three healthy subjects using the dual refocusing echo mode (DREAM) technique (1,2). The DREAM scan was acquired using an ECG-triggered Cartesian 2D single-shot trajectory with 8 averages in the same mid-ventricular location as the Multimapping slice. Two 60° pulse, spaced 5 ms apart, preceded the image acquisition, which consisted of dual echo readouts to separate the free induction decay and the stimulated echo components. Other imaging parameters were: field-of-view = 350×350 mm, spatial resolution = 4×4 mm, slice thickness = 10 mm, flip angle

= 20°, bandwidth = 721 Hz/pixel, $TR/TE_1/TE_2$ = 5.3/1.6/3.1 ms, SENSE factor = 2. The acquisition time for each single-shot image was 200 ms, and triggered to coincide with the mid-diastolic rest period. DREAM $B_1$ maps were generated using a vendor-provided algorithm (2). Mean $B_1$ was measured in the Multimapping and DREAM maps according to the 6 AHA segments, as well as across the entire slice.

Example Multimapping and DREAM $B_1$ maps from one healthy subject are shown in Figure S5.

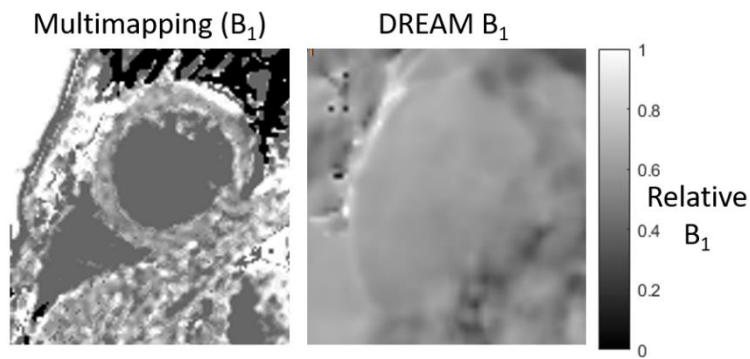

*Figure S5. Example Multimapping and dual refocusing echo mode (DREAM) $B_1$ maps from one healthy subject. Note, $B_1$ cannot be accurately estimated in the blood pool for Multimapping due to significant through-plane blood flow.*

The $B_1$ measurements for all segments and subjects are shown in Figure S6. Multimapping $B_1$ is in good agreement with DREAM and provide maps within 0.1 relative $B_1$ for 14 out of 18 (78%) analyzed segments. Differences in septal $B_1$ are all within 0.08 relative $B_1$ suggesting this is a preferable location to estimate $B_1$. Finally, mean intra-subject relative $B_1$ variability, defined as the mean standard deviation between segments for the three subjects, was 0.046 for DREAM, indicating that $B_1$ is homogeneous across the slice. These findings justify the use of a septal measurement to estimate $B_1$ for the entire slice for Multimapping.

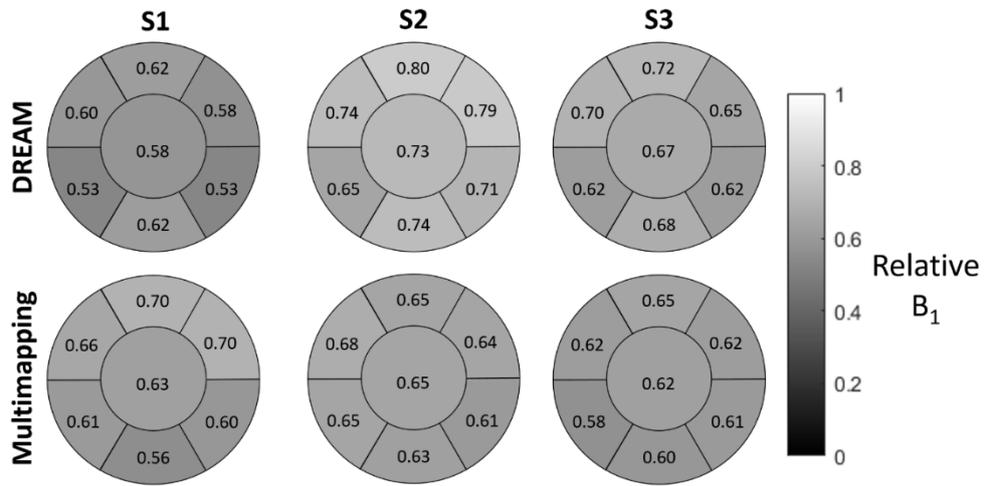

*Figure S6. Multimapping and DREAM relative mean B₁ for the three subjects (S1-3).*

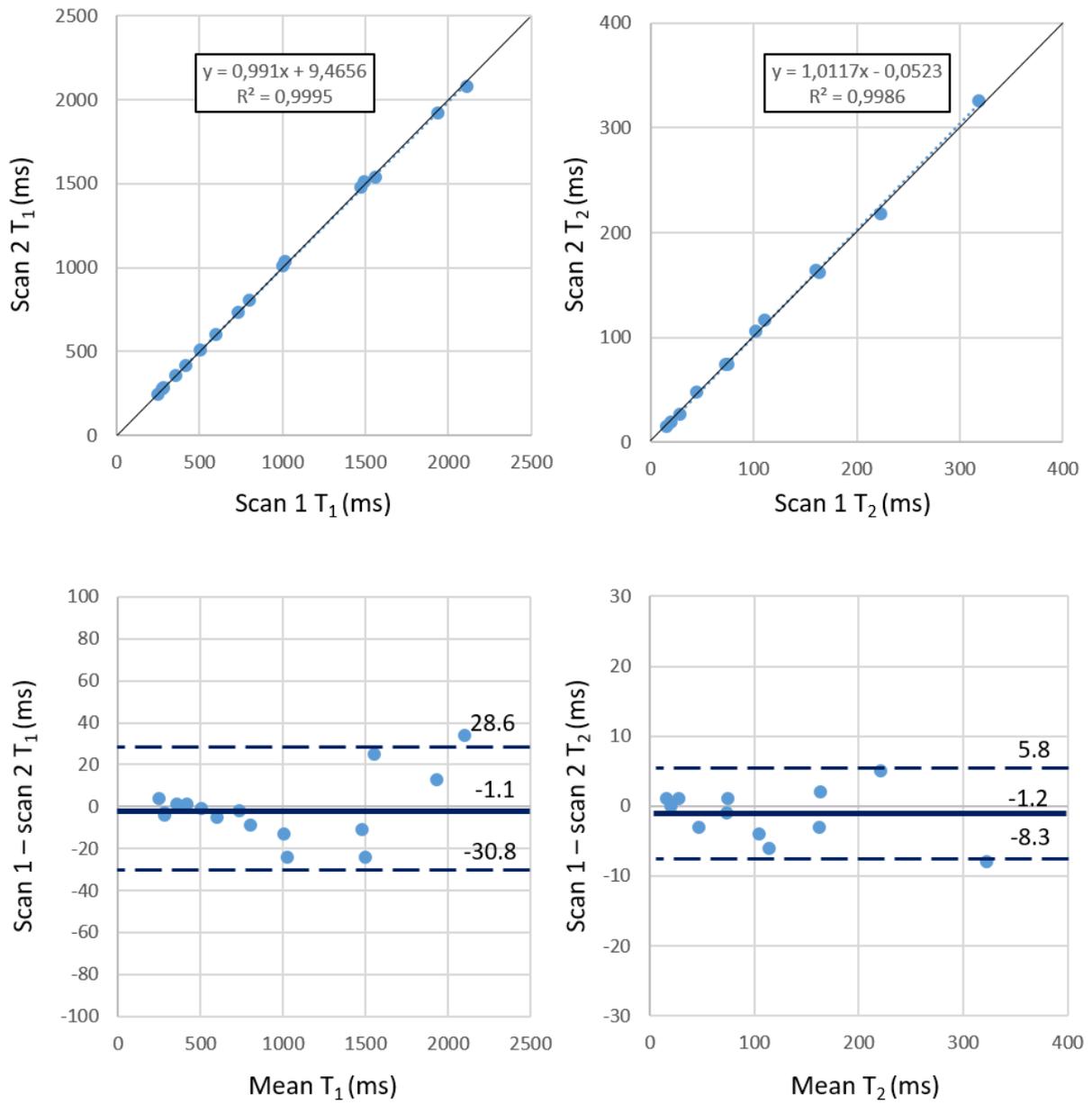

*Figure S7. Correlation and Bland-Altman plots for the phantom $T_1$ and $T_2$ repeatability experiments.*

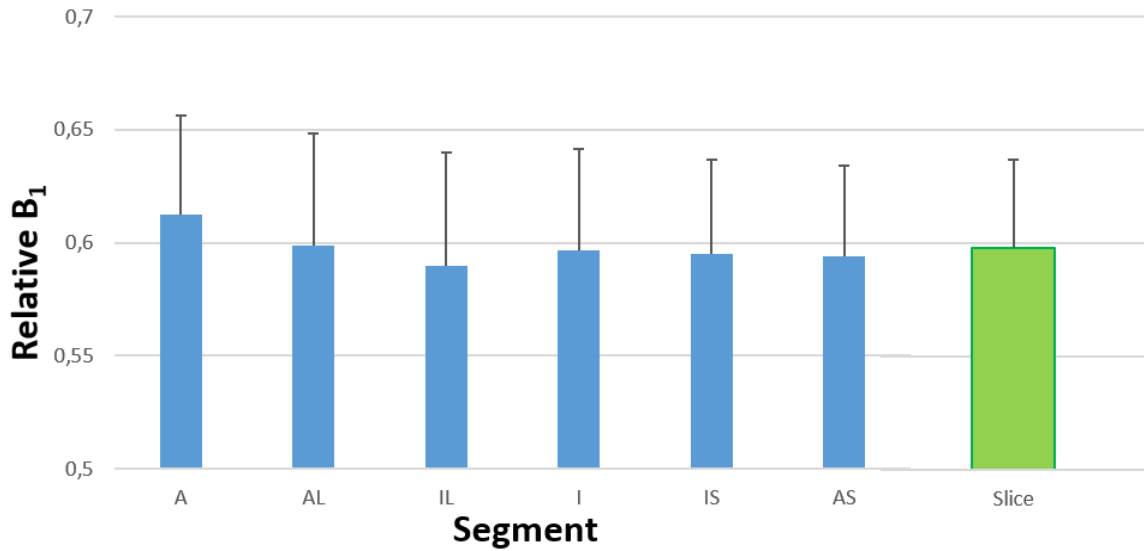

*Figure S8. Mean relative $B_1$ for all 16 healthy subjects using Multimapping for the 6 mid-ventricular AHA segments (A=anterior; AL=anterolateral; IL=inferolateral; I=inferior; IS=inferoseptal; AS=anteroseptal), and averaged across the slice.*

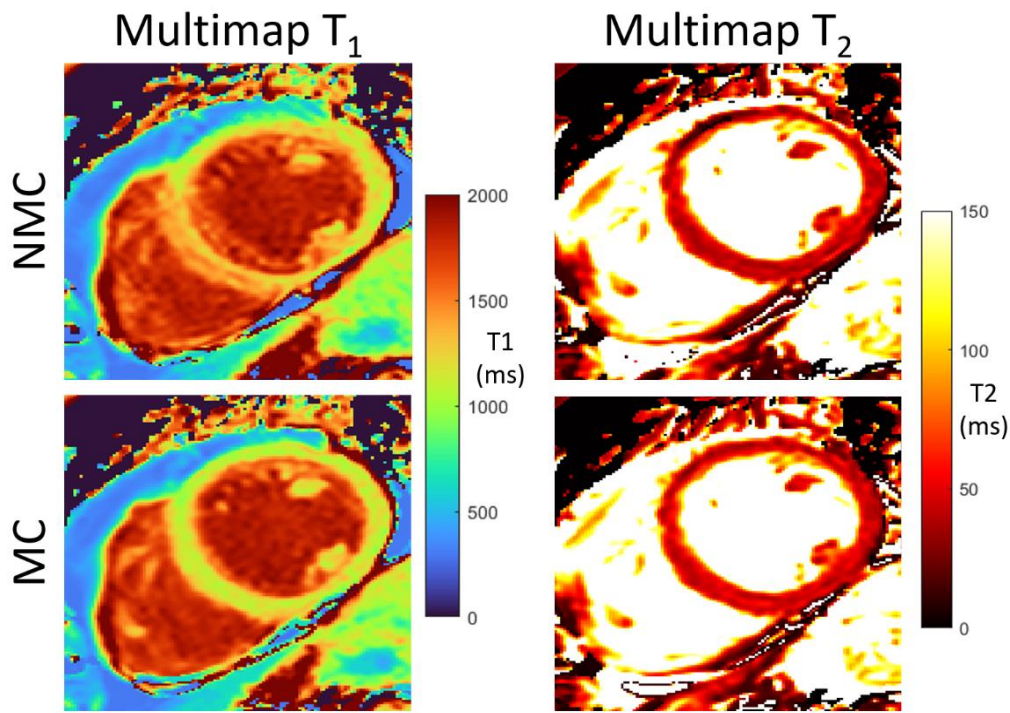

*Figure S9. Multimaps acquired in one healthy subject with substantial respiratory motion, generated with no motion correction (NMC), and with motion correction (MC) using nonlinear image registration. Improved sharpness can be noted in the septum of the motion corrected $T_1$ maps, while the removal of a motion-induced artifact in the anterior segment can be seen in the motion corrected T2 maps.*

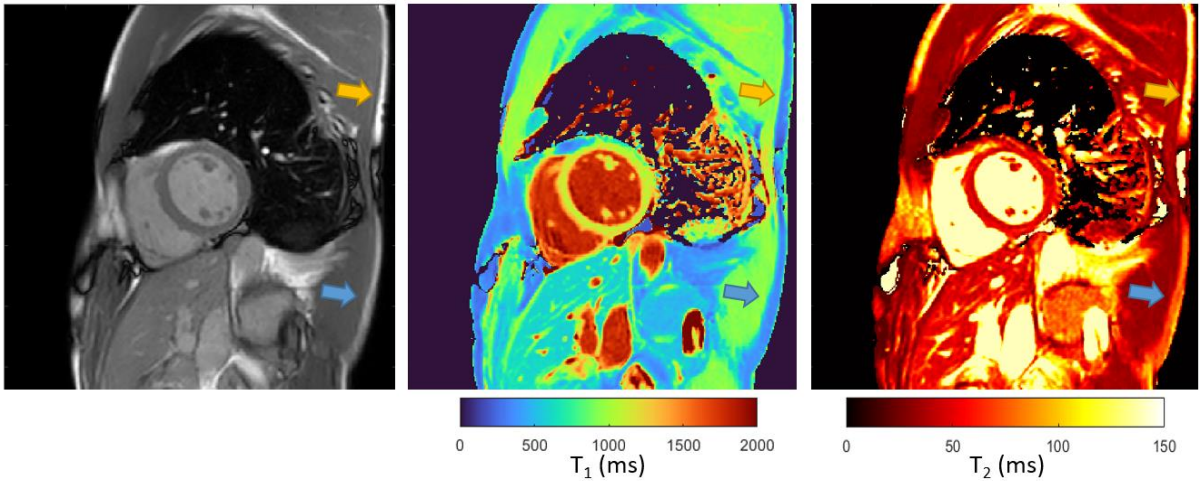

*Figure S10. Multimapping source image, $T_1$ and $T_2$ map from one healthy subject showing differences in quantification of subcutaneous fat (arrows), which is particularly noticeable for $T_2$. $T_1$ in the superior region-of-interest (orange arrow) was 298 ms, while $T_1$ in the inferior region-of-interest (blue arrow) was 320 ms. $T_2$ in the corresponding regions were 112 ms and 78 ms, respectively.*

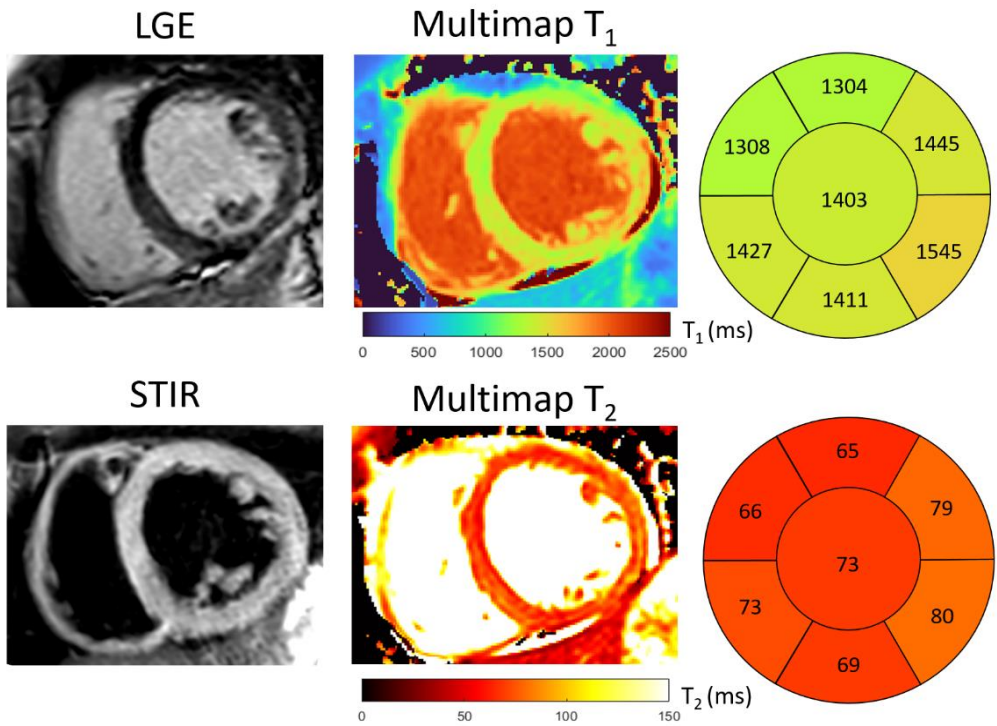

*Figure S11. $T_1$ and $T_2$ Multimaps, late gadolinium enhancement (LGE) and short tau inversion recovery (STIR) in the 41-year-old patient with acute myocarditis. Significant increases in $T_1$ and $T_2$ were seen throughout the myocardium, particularly in the inferolateral segment as shown in the bullseye plots on*

the right. The LGE images indicate gadolinium uptake in the myocardium, although it is relatively diffuse, while the STIR (image intensity ratio = 2.0) is borderline suggestive of edema.

Table S1. Normalized mean root square error (NMRSE) for $T_1$ and $T_2$ in the phantom scans.

| Heart rate (bpm) | $T_1$ NMRSE (%) mean [min max] | $T_2$ NMRSE (%) mean [min max] |
|---|---|---|
| 40 | 4.2 [0.5 11.5] | 6.2 [0 15.2] |
| 60 | 2.9 [0 7.6] | 6 [0 15.4] |
| 80 | 2.6 [0 7.1] | 10 [2.5 20.7] |
| 100 | 2.6 [0.3 6.4] | 7.7 [0 18.5] |
| 120 | 2.4 [0 6.4] | 11.9 [2.5 32.9] |